# Ethereum ECCPoW

HYOUNGSUNG KIM[1], JEHYUK JANG[2], SANGJUN PARK[3], AND HEUNG-NO LEE[1], (Senior Member, IEEE)

[1,2]School of Electrical Engineering and Computer Science, Gwangju Institute of Science and Technology, Gwangju 61005, South Korea
[3]Electronics and Telecommunications Research Institute (ETRI), South Korea

Corresponding author: Heung-No Lee (heungno@gist.ac.kr).

This work was partially supported by the National Research Foundation of Korea (NRF) grant funded by the Korean government (MSIP) (NRF-2018R1A2A1A19018665). This research was partially funded by Institute of Information & Communications Technology Planning& Evaluation, grant number 2020-0-00958.

**ABSTRACT** The error-correction code based proof-of-work (ECCPoW) algorithm is based on a low-density parity-check (LDPC) code. The ECCPoW is possible to impair ASIC with its time-varying capability of the parameters of LDPC code. Previous researches on the ECCPoW algorithm have presented its theory and implementation on Bitcoin. But they do not discuss how stable the block generation time is. A finite mean block generation time (BGT) and none heavy-tail BGT distribution are the ones of the focus in this study. In the ECCPoW algorithm, BGT may show a long-tailed distribution due to time-varying cryptographic puzzles. Thus, it is of interest to see if the BGT distribution is not heavy-tailed and if it shows a finite mean. If the distribution is heavy-tailed, then confirmation of a transaction cannot be guaranteed. We present implementation, simulation, and validation of ECCPoW Ethereum. In implementation, we explain how the ECCPoW algorithm is integrated into Ethereum 1.0 as a new consensus algorithm. In the simulation, we perform a multinode simulation to show that the ECCPoW Ethereum works well with automatic difficulty change. In the validation, we present the statistical results of the two-sample Anderson-Darling test to show that the distribution of BGT satisfies the necessary condition of the exponential distribution. Our implementation is downloadable at https://github.com/cryptoecc/ETH-ECC.

## I. INTRODUCTION

Blockchain is a peer-to-peer (P2P) network that consists of trustless nodes. In a reliable P2P network, no peers would intentionally send wrong information to others. In contrast, in an unreliable P2P network (e.g., a group of trustless nodes), the possibility that some peers may send false information to others should be considered. For example, a node may spread wrong or forged information to others. To address these issues in an unreliable network, Nakamoto has proposed the ideas of using blocks and chaining of blocks with a novel consensus algorithm [1].

In a blockchain, one of the peers propagates a new block containing transactions to other peers. Peers validate the received block and link it to the previous block when there is no problem in the received block. A consensus algorithm accomplishes this process. If one of the peers has sent false information to others, such information is detected by the consensus algorithm as there is no collusion among the peers. A generated block contains information about previous blocks; thus, if someone wants to change one block in a chain, all previous blocks of the changing block must change. Therefore, unless the network is centralized within a particular group, sending forged information about previous blocks to new peers is impossible. Therefore, to prevent collusion, the unreliable network should avoid centralization.

Nakamoto has proposed a proof-of-work (PoW) system for a consensus algorithm. In the PoW system, peers repeat a type of work to solve a cryptographic puzzle using a hash function (e.g., SHA256 [1], Keccak [2]). When a peer successfully solves a cryptographic puzzle, the peer generates a block. Additionally, the node gets an incentive as a reward for the work done. In an ideal PoW system, anyone can join to work and take as much incentive as they can get the reward for the completed work. However, with an increase in the price of reward, attempts have been made to centralize the network to monopolize incentives.

Centralization is a phenomenon occurring in PoW based blockchain networks. In blockchains utilizing PoW as a consensus algorithm, an oligarchy of miners who possess overwhelming portion of computation resources can monopolize the chance to generate blocks. Such centralization negatively impacts the credibility of a blockchain. For example, in a centralized network, a group of dominant nodes can selectively filter out some transactions belonging to others for their own benefits. As far as new nodes are concerned, it will be difficult for them to give trust and join the network in fear of possible unfair treatment [3], [4].

The emergence of application-specific integrated circuits (ASIC) has accelerated the centralization of PoW. As more nodes use ASICs in generating blocks, block generation requires more computations. Thus, it has become hard to generate blocks using general-purpose units, such as a central processing unit (CPU) and a graphics processing unit (GPU). As a result, a few groups equipped with powerful ASICs have surfaced and centralized the blockchain networks. To avoid centralization, researchers have proposed the use of ASIC-resistant PoW (e.g., Ethash [2], X11 [12], Random X [24]) and alternative consensus algorithms (e.g., proof-of-stake, delegated-proof-of-stake, or Byzantium fault tolerance [25]). The networks of alternative algorithms have presented less decentralization effects than the network of ASIC-resistant PoW does [25]. Specifically, in alternative algorithms, only limited participants can generate blocks; but, ASIC-resistant PoW has no limit of participants. Thus, ASIC-resistant PoW presents a better-decentralized network than alternative algorithms.

For an ASIC-resistant PoW, an error-correction code based proof-of-work (ECCPoW) algorithm was proposed [6], [7]. In ECCPoW algorithms, a hash value of a previous block generates a varying parity check matrix (PCM) for error-correction. This varying PCM works as a cryptographic puzzle in ECCPoW. These time-varying cryptographic puzzles make ECCPoW ASIC-resistant. It is possible to implementing an ASIC for a specific cryptographic puzzle. In ECCPoW, every newly created puzzle differs from all the previously created puzzles. As a result, if there is an ASIC for ECCPoW, such an ASIC must cover a wide range of cryptographic puzzle generation systems. Such a system, however, would incur huge chip space and cost [10], [11].

In [7], the authors have reported that the time-varying puzzle system may generate large block generation times (BGT), i.e., outliers, for ECCPoW implemented on Bitcoin. If the outliers occur frequently enough, it is of our interest in this paper to see, the distribution of BGT may show a heavy-tailed distribution with a none finite mean [15], [26]. As a result, the definition of [6] that BGT has a finite mean needs to be challenged. Previous works on ECCPoW [6], [7] did not include real-world experiments extensive enough to conclude that BGT has a finite mean. If BGT does not have a finite mean, ECCPoW cannot be used as an Ethereum consensus algorithm. Therefore, in this paper, we aim to study the distribution of BGT of the ECCPoW implemented on Ethereum (ETH-ECC). Our experimental results show that the BGT distribution is not heavy-tailed and has a finite mean.

The contributions of our work are as follows:

- We show how ECCPoW is implemented on Ethereum.

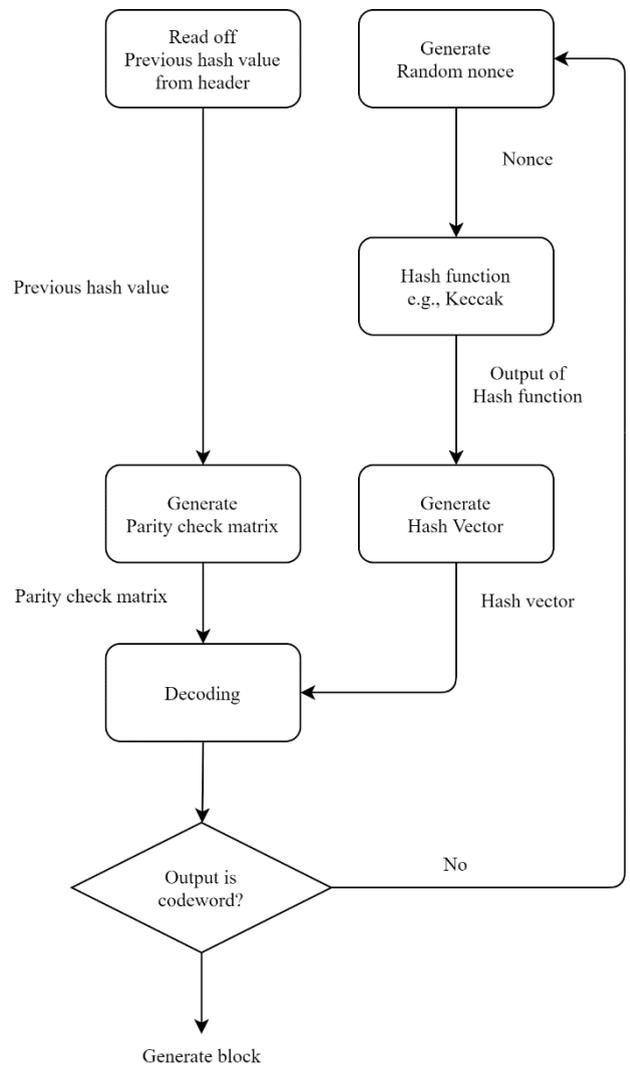

FIGURE 1. Flow chart of ECCPoW Ethereum. Every miner who generates blocks can make a parity check matrix using a previous hash value. A generated nonce becomes an input of a hash function. A hash vector used for decoding can be generated using the output of a hash function. If decoding is successful, the block is generated; otherwise, a miner generates a new nonce to make a new hash vector for decoding.

- We present a method to control the difficulty in ETH-ECC and report the results of automatic difficulty change with real-world experiments of ETH-ECC.
- We present a goodness-of-fit result using the Anderson–Darling (AD) test for distribution validation and discuss the necessary condition that the BGT distribution of ETH-ECC follows the exponential distribution.

The remainder of this paper is organized as follows. Section II provides a background of the requirements of an ASIC-resistant PoW. Section III demonstrates the implementation of the ETH-ECC. Section IV discusses the formulation of the problem. Section V provides the experimental result of the implementation of the ETH-ECC. Finally, Section VI summarizes our work and concludes the paper.

## II. Background

We introduce three approaches that can use to avoid centralization problems in PoW. One is an intentional bottleneck between an arithmetic logic unit (ALU) and memory, which is used by Ethash of Ethereum [2], [5]. It is also termed a memory-hard technique. Another one is the *high complexity of ASIC design* utilized by Dash [12], Raven [13], and in our method, ECCPoW. The third one is *hybrid methods* of two methods; Random X of Monero utilizes *hybrid methods* [24].

### A. INTENTIONAL BOTTLENECK

The most known PoW of the intentional bottleneck is Ethash of Ethereum [2], [5]. This method uses the difference between the throughput of ALU and the bandwidth of the memory. If there is a bottleneck between the ALU and memory, it is impossible to fully utilize the throughput of ALU. Specifically, if a miner who wants to generate a block must get data from memory to generate a block, the number of block generation attempts depends on the memory bandwidth. The Ethash uses a directed acyclic graph (DAG), which is a set of randomly generated data for the bottleneck. The DAG is a huge dataset that cannot be stored in cache memory; therefore, DAG is stored in memory. To generate a block using Ethash, a miner must mix a part of DAG that is stored in the memory. Owing to this procedure, the miner cannot avoid the bottleneck, which is derived by the memory bandwidth. This method has been ASIC-resistant for a long time; however, Bitmain released ASIC for Ethash in 2018.

### B. HIGH COMPLEXITY OF ASIC DESIGN

The high complexity of ASIC design forces ASIC to be less efficient. For example, if ASIC is less efficient than a general-purpose unit such as CPU or GPU, there is no reason to design ASIC. X11 of Dash [12] and X16R of Raven [13] utilize this method. Unlike PoW of Bitcoin, which uses only one hash function (SHA-256), X11 uses 11 hash functions consecutively: BLAKE, BMW, Grosetl, JH, Keccak, Skein, Luffa, Cubehash, SHAvite-3, SIMD, and ECHO. The BLAKE, which is the first hash function of X11, uses a block header with nonce as inputs; and its output becomes the input of the next hash function. Similarly, the next hash function uses the output of the previous hash function. This procedure is repeated until a result is obtained for the last hash function ECHO. Using the result of last hash function, miners determine whether they find a valid nonce.

Designing an ASIC for X11 was expensive; therefore, X11 was ASIC resistant. However, Bitmain released an ASIC for X11 in 2016. There are a few PoW algorithms that extend X11 (e.g., X13, X14, and X15); however, the ASICs for these have been released. X16R of Raven is an extended version of X11 of Dash. In X16R, unlike the previous extension of X11, the sequence of 16 hash functions is randomly changed. Therefore, it is costly to design an ASIC for X16R. However, T. Black, who designed X16R, mentioned that there is some evidence that ASICs for X16R exist [23]. Our ECCPoW also uses *high complexity of ASIC design* method for ASIC resistance. However, unlike previous algorithms, ECCPoW can make ASIC powerless despite the release of ASIC. We explain this detail in Section III.

### C. HYBRID METHODS

Random X of Monero combines the above two methods. Random X uses memory-hard techniques for the bottleneck with random code execution; Random X is optimized for CPU mining [24]. In [24], they mentioned that it is possible to perform mining using field programmable gate array (FPGA); however, it will be much less efficient than CPU mining. It implies that it is possible to make efficient mining hardware when the cost of developing chip sets is low relative to the mining reward. With the proposed ECCPoW, attempts to develop an efficient mining hardware can be made when the reward-to-cost ratio gets higher. However, such attempts can be evaded easily since the parameters of ECCPoW can be easily changed, such as increasing the length of code and the code rate. The next section describes more on the ASCI-resistance characteristic of ECCPoW.

## III. ECCPoW Implemented on Ethereum

In this Section, we aim to briefly introduce ECCPoW and present how ECCPoW has been implemented on Ethereum using Fig. 1. In addition, we present how the difficulty level of ETH-ECC can be controlled automatically.

### A. OVERVIEW OF THE ECCPoW

In a blockchain employing the PoW consensus algorithm, a node solves cryptographic puzzles to publish a block. For a given puzzle, the node who solves the puzzle first gets the authority to publish a block. For example, in the PoW of Bitcoin, the first node that finds a specific output of the Secure hash algorithm (SHA) gets the authority to publish a block. The PoW of Ethereum uses Keccak instead of SHA. The ECCPoW algorithm proposed in [6] is a PoW consensus algorithm that utilizes error-correction code, which is made of the low-density parity-check (LDPC) code [8], as a cryptographic puzzle. The ECCPoW algorithm consists of a pseudo-random puzzle generator (PRPG) and an ECC puzzle solver. Fig. 1 presents the flow chart of the ECCPoW algorithm. For every block, the PRPG generates a new pseudo-random LDPC matrix. LDPC matrix is distinct from the other previously generated matrices. Such a pseudo-random LDPC matrix takes the role of issuing an independently announced cryptographic puzzle. The ECC puzzle solver uses the LDPC decoder to solve the given announced puzzle. Specifically, to publish a block, a node is required to run through input header until the LDPC decoder hits a satisfying result; say, the output of the decoder is an LDPC codeword (with a certain Hamming weight). In the next subsection, we discuss ECCPoW implementation on Ethereum with the flow chart presented in Fig. 1.

### B. ECCPoW ON ETHEREUM

In this subsection, we present how the error-correction process is applied to ETH-ECC using Fig. 1.

$$C := \{\mathbf{c} \mid \mathbf{Hc} = \mathbf{0} \cap \mathbf{c} \in \{0,1\}^{n \times 1}\} \quad (1)$$

when a parity-check matrix (PCM) **H** is given, a code **c**, which satisfies (1) is referred to as an LDPC code. The goal of the ECCPoW algorithm is to find an LDPC code **c** using the PCM **H**, which is derived by PRPG, and a hash vector **r**, which is obtained by the ECC puzzle solver. For the PRPG, we employ the previous hash value; the previous hash value, known as the parent hash in the Ethereum block header, randomly generates a PCM. Specifically, we use Gallagher's method to make random PCM [9]; we use the previous hash value as a seed of randomness. Thus, PCM is changed every block; because of the same seed, every node uses the same PCM until a block is generated [6].

### 1) ECC puzzle solver on ECCPoW Ethereum

Here, we introduce a process of ECC puzzle solver in ETH-ECC. Our definitions are based on [6]. The equations below follow the right-hand side of Fig. 1.

**Definition 1**. Hash vector **r** in which the size of $n$ can be obtained as follows:

$$s_1 := Keccak(nonce) \in \{0,1\}^{256} \quad (2)$$

where *Keccak* denotes the hash function applied in Ethash of Ethereum [5]. We use the same way of Ethereum to generate a *nonce*. Furthermore, for a longer length of hash vector, we use $s_u := Keccak(s_1) \in \{0,1\}^{256}$ with $u = 2, 3, \ldots, l+1$. We slice or concatenate the result of *Keccak* to generate a flexible length hash vector **r**:

$$\mathbf{r} := \begin{cases} s_1[1:n] & \text{if } n \leq 256 \\ [s_1 \cdots s_l \; s_{l+1}[1:j]] & \text{if } n > 256 \end{cases} \quad (3)$$

where $l = \lfloor n/256 \rfloor$ and $j = n - 256 \times l$. For example, when $n$ is less than 256, **r** gets the same length as $n$; and when $n$ is not less than 256, **r** concatenates the results of *Keccak*. These flexible length hash vector is utilized for ASIC-resistance.

### 2) Proof-of-Work of the LDPC decoder

The goal of the LDPC decoder is to find a hash vector **c** that satisfies **Hc** = 0. The below definition explains the decoding presented in Fig. 1.

**Definition 2.** When PCM **H**, which is the size of $m \times n$, and hash vector **r**, which is the size of $n$, are given, the LDPC decoder uses **H** and **r** as inputs and obtains output **c** using the message-passing algorithm [6], [14]. When **c** satisfies (1), **c** becomes an LDPC code, and the miner completes LDPC decoding.

$$D_{np} : \{\mathbf{r}, \mathbf{H}\} \mapsto \mathbf{c} \in \{0,1\}^{n \times 1} \quad (4)$$

A PCM **H** is randomly generated, but all miners use the same previous hash value, which is derived from the previous block. Therefore, it is impossible to predict the next PCM to mine a block in advance. In the PoW of Ethereum, miners change a nonce when they got a wrong output. We follow the same way as that used by Ethereum to obtain a hash value from *Keccak* with a *nonce*, but ETH-ECC uses one more step (3) to generate a hash vector for decoding. When the code derived by (4) does not satisfy (1), the miner generates a new *nonce* and repeats all the steps.

Our method is based on the *high complexity of ASIC design* in Section II for an ASIC-resistant PoW. However, unlike the mentioned method in Section II, ECCPoW generates varying cryptographic puzzles for *high complexity*. Specifically, ECCPoW utilizes two factors for *high complexity*: flexible length LDPC code **c**, randomly generated PCM **H**. ASICs can be released for the $n$ length of code. However, extending the length of code (e.g., $n+1$) makes ASICs powerless. Furthermore, in [10], [11], it has been proven that implementing an ASIC that can handle variable PCM is expensive and occupies a lot of space. If developing an ASIC costs more than buying a CPU or GPU, there is no incentive to make an ASIC. In other words, the ECCPoW algorithm is ASIC resistant as implementing an ASIC that can handle various lengths of changing codes and randomly generated PCMs is very inefficient.

### C. DIFFICULTY CONTROL OF ETH-ECC

In this subsection, we demonstrate the implementation of difficulty control of ETH-ECC. Bitcoin [1] and Ethereum [2] have different difficulty control methods. In Bitcoin, the Bitcoin network changes the difficulty every 2016 block; the desired block generation time is 10 minutes for a block. If miners generate a block every 10 minutes, generating 2016 blocks takes precisely two weeks. Thus, if generating 2016 blocks takes more than two weeks, the difficulty decrease; on the contrary case, the difficulty increase. Unlike Bitcoin, the Ethereum network changes the difficulty every block. Ethereum network allows for generating a block between nine seconds and 18 seconds. If a block is generated within nine seconds, then the difficulty increase. If it exceeds 18 seconds, then the difficulty decrease. Because of this difference between Bitcoin and Ethereum, ECCPoW based Bitcoin(BIT-ECC) and ETH-ECC also have different difficulty control methods. Thus, ETH-ECC cannot utilize BIT-ECC's method. Because of the need for a new method, we demonstrate the implementation of difficulty control of ETH-ECC with a difference from Ethereum's method.

Ethereum utilizes the number of attempts to generate a block per second, termed as hash rate, and a probability of block generation. Similarly, ETH-ECC utilizes the hash rate; but ETH-ECC considers a probability of decoding success. In [5], the difficulty of Ethereum is defined by the probability of block generation. The difficulty follows:

$$n \leq \frac{2^{256}}{Diff} \quad (5)$$

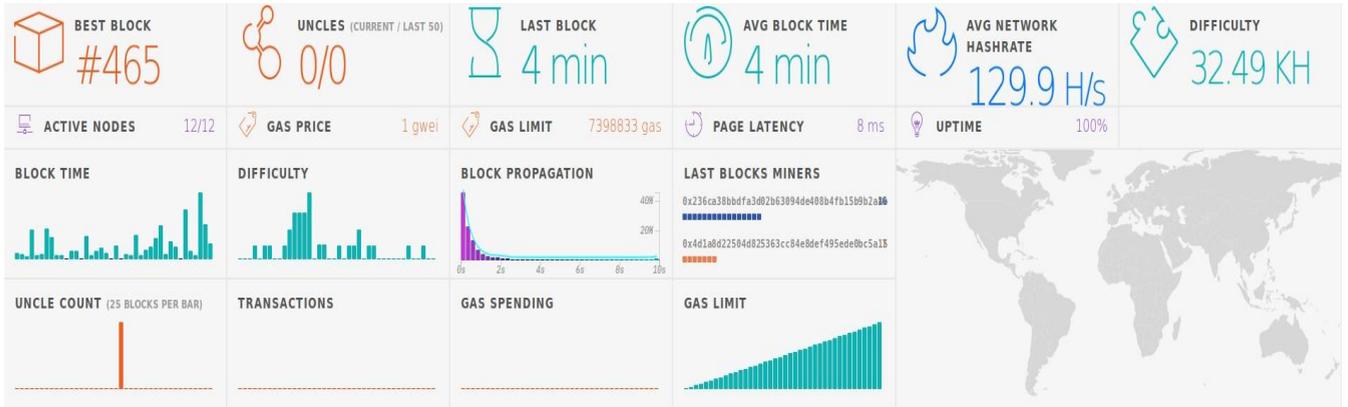

**FIGURE 2.** This figure shows the results of the simulation of ECCPoW Ethereum on Amazon Web Services (AWS). Twelve nodes are used in the simulation. The two nodes are *bootnodes* that help connect the nodes, and the other 10 nodes are *sealnodes* that participate in the block generation. We use the m5.xlarge of AWS EC2 for the simulation. In the charts, *BLOCK TIME* shows the block generation times for the last 40 blocks, and *DIFFICULTY* shows the difficulty levels of the last 40 blocks. *BLOCK PROPAGATION* shows the percentage of the block propagation time corresponding to time.

It indicates that

$$Diff \leq \frac{2^{256}}{n} \quad (6)$$

where $n$ denotes the result of PoW, and *Diff* denotes the difficulty of Ethereum. Thus, (6) means that when the difficulty increases, the number of $n$ that satisfies (6) decreases. Furthermore, we can consider that the reciprocal of difficulty is a probability of block generation. Ethereum utilizes this probability and hash rate to control block generation time. For example, without replacement, when the probability of block generation is 1/150 and hash rate is 10 hash per second, brute force takes 15 seconds. If the hash rate increases, such as 20 hash per second, Ethereum's method adjust the probability of block generation time to 1/300. Thus brute force takes 15 seconds even though the hash rate increase.

For ECCPoW, if we can calculate a probability of decoding success, it is possible to control difficulty similar to the process in Ethereum. Thus, it is important to know the probability of a successful LDPC decoding according to the LDPC parameter. To test the difficulty change using the BGT, we use the pseudo-probability of a successful LDCP decoding according to the parameters [7]. Namely, ETH-ECC utilizes the probability of decoding success and hash rate to control difficulty. For example, without replacement, when the probability of decoding success is 1/150, and the hash rate is 10 hash per second, it takes 15 seconds like the above example of Ethereum's method. However, unlike Ethereum, when the hash rate increase, ETH-ECC tunes parameters of LDPC to adjust the probability of decoding success. By tuning parameters, ECCPoW achieves both difficulty control and ASIC-resistance. The parameters can be found at https://github.com/cryptoecc/ETH-ECC/blob/master/consensus/eccpow/LDPCDifficulty_utils.go#L65. In Fig. 2, the difficulty of the ETH-ECC is 32.49 KH. It is indicating that the probability of block generation is 1 of 32,490 hash.

### IV. Problem Formulation
In PoW, there is a case that nodes generate blocks at the same time. Bitcoin allows only one block; Ethereum allows three blocks to generate at the same time. However, in Ethereum, only one block can be canonical; the other blocks cannot. Blocks that cannot be the canonical is called an uncle block. In Ethereum, nodes rollback transactions of uncle blocks [5]. Therefore, the transaction's participants must wait block confirmation to prevent the rollback. That is to say, in blockchains utilizing PoW, the BGT must have a finite mean for the block confirmation time. For example, if the BGT has a none finite mean, we cannot determine how long we must wait for the confirmation of transactions. Therefore, to apply the ECCPoW algorithm in a real network, the BGT must have a finite mean.

In [6], the authors present the definition of the block generation of the ECCPoW algorithm using a hash rate with a geometric distribution. Namely, they assumed that nodes generate a block within specific block generation attempts. However, if the BGT has a none finite mean, there is no guarantee that nodes generate a block within specific attempts. In [7], the authors present a practical experiment using the ECCPoW algorithm. However, they only mentioned that the BGT of ECCPoW is "unstable". Namely, they mentioned that BGT of ECCPoW has outliers; but they did not present a discussion about BGT. Thus, in the paper, we present a discussion about BGT. Specifically, our experimental result presents evidence that exponential distribution describes the distribution of BGT of ECCPoW.

### V. EXPERIMENT ON ETH-ECC
In this section, we conduct experiments using ETH-ECC. First, we simulate the difficulty change using multinode networks. Second, we conduct a goodness-of-fit experiment using the Anderson-darling (AD) test [16], [17], [18] to discuss the distribution of the BGT with fixed difficulty.

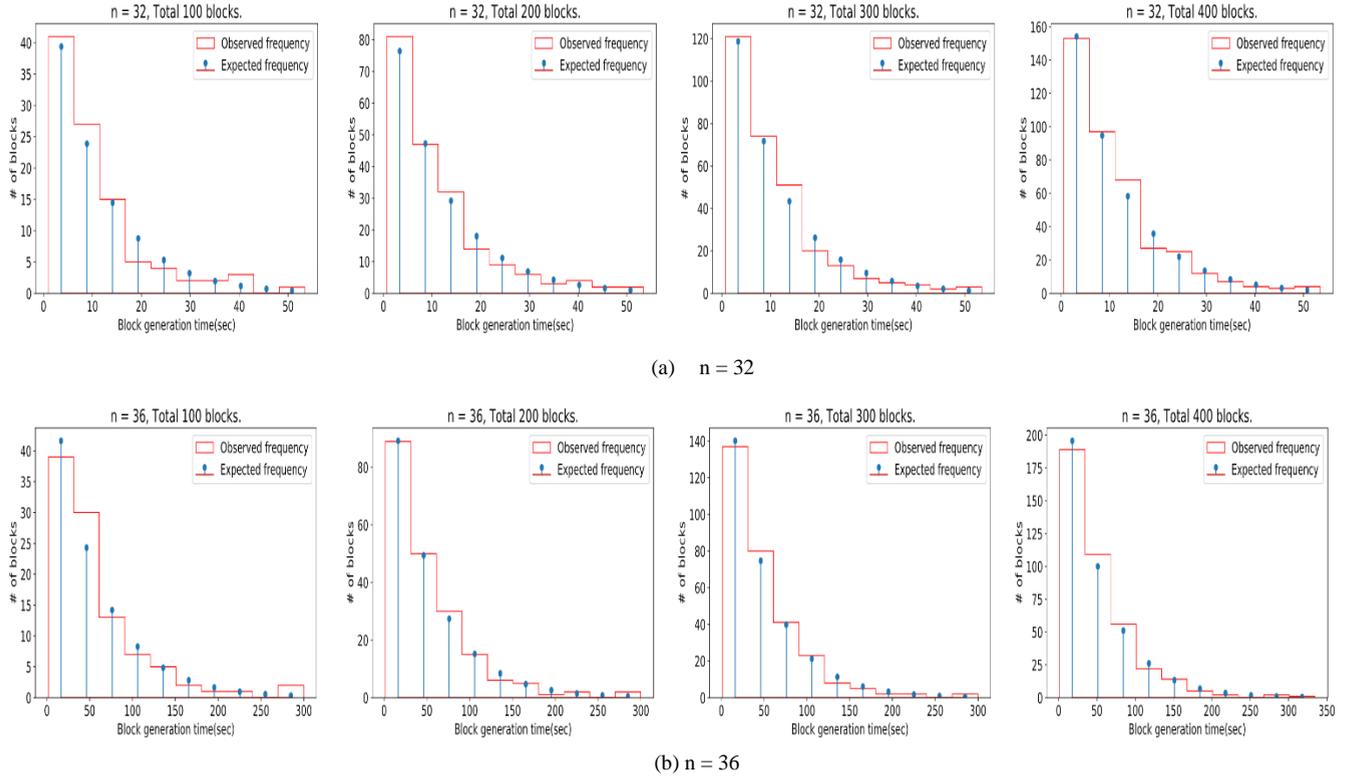

**FIGURE 3.** The numbers of all the blocks are 100, 200, 300, and 400. The expected frequency is calculated using the exponential distribution derived from the mean of the observed block generation time.

### A. SIMULATION OF THE DIFFICULTY CHANGE

We simulate the difficulty change employing Amazon Web Services (AWS) using 12 nodes. Two nodes are *bootnodes* that help connect the nodes, and the other 10 nodes are *sealnodes* that participate in the block generation. In the charts presented in Fig. 2, *BLOCK TIME* presents the BGT of the last 40 blocks, and *DIFFICULTY* shows the difficulty of the last 40 generated blocks. *BLOCK TIME* and *DIFFICULTY* show that because of the large standard deviation, the block is generated slow despite the low level of difficulty, as already mentioned in [7]; in the next subsection, we discuss the BGT. In the charts presented in Fig. 2, *LAST BLOCK* shows the BGT of the previous block, and *AVG BLOCK TIME* shows the average of the BGT. Moreover, *AVG NETWORK HASHRATE* shows the average hash rate of all miners. *BLOCK PROPAGATION* shows the block propagation time from a miner who generated a block to other miners. We used two different regions: Seoul and US East for *sealnodes*. Specifically, 3 of the 10 *sealnodes* are in the US East region, whereas the rest are in the Seoul region. *BLOCK PROPAGATION* also shows the percentage of blocks which are propagated corresponding times. *BLOCK PROPAGATION* indicates that propagation of almost blocks takes less than 2 seconds to propagate between the Seoul and US East regions. Block propagation follows the same method as that of Ethereum.

### B. STABILTY OF THE BLOCK GENERATION TIME

Fig. 2 demonstrates the need to check if varying puzzles might make outliers. Namely, in *BLOCK TIME* and *DIFFICULTY* of Fig. 2, a slow block generations are observed despite the low level of difficulty. In other words, the observation of BGT shows outliers. If the outliers are not controllable, outliers make the distribution of BGT have none finite mean similar to the heavy-tailed distribution. None finite mean cannot guarantee the confirmation of transactions. Thus, to achieve a stable BGT that can guarantee the confirmation of transactions, the BGT must have a finite mean.

We obtain the BGT of ECCPoW Ethereum with a fixed difficulty to observe what kind of distribution with a finite mean the BGT follows. Specifically, if BGT follows exponential, it has a finite mean. However, if the BGT follows a heavy-tailed distribution, it has a none finite mean [15]. Thus, through the goodness-of-fit, we aimed to discuss what type of distribution the BGT follows. For the goodness-of-fit, we set a null hypothesis $H_0$ and alternative hypothesis $H_A$:

$H_0$ : BGT has the exponential distribution

$H_A$ : BGT does not have the exponential distribution

For the goodness-of-fit, we use the AD test [16], [17], [18]. There are other tests available for the goodness-of-fit such as the chi-squared test [19], Kolmogorov–Smirnov test [20], and AD test [16]. The chi-squared test has a restrictive assumption that all the expected frequencies should be five or more [21]. But, there is no guarantee that our samples achieve this assumption. If we collect more samples, the chi-squared test possibly uses. However, the *p*-values used to validate the hypotheses are affected by the number of samples. When the number of samples increased in the chi-squared test, the *p*-values tend to decrease. Therefore, the assumption of the chi-squared test is not appropriate for verifying our distributions. The Kolmogorov–Smirnov test does not have an issue with adequacy on sample size. But it is sensitive more to the center of the distribution rather than the tail [22]. To cover all possibilities, we must consider verifying the tail of the distribution. Therefore, we have chosen to use the AD test [16], which gives more weight to the tail compared with the Kolmogorov–Smirnov test.

### C. Anderson-Darling Tests

In this subsection, we aim to discuss the Anderson-Darling test and verify its usage using test examples. The AD test is used to verify if a sample follows a specific distribution. We discuss one-sample and two-sample AD tests. In our work, we use the two-sample AD test; but to present our contribution clearly, we briefly introduce the one-sample AD test first.

#### 1) One-sample AD test

The one-sample AD test is suitable to verify a hypothesis that a sample set comes from a population. The one-sample AD test is as follows. When the cumulative distribution function (CDF) of the population distribution is $F(x)$ and CDF of the empirical distribution is $F_M(x)$, the one-sample AD test [18] is used as follows:

$$A_M^2 = M \int_{-\infty}^{\infty} \left(F_M(x) - F(x)\right)^2 w(x) dF(x) \quad (7)$$

and

$$w(x) = [F(x)(1 - F(x))]^{-1} \quad (8)$$

where $M$ denotes the number of samples, and $A_M^2$ denotes the results of the one-sample AD test. Intuitively, in (7), if $F_M(x) - F(x)$ is 0 for all $x$, $A_M^2$ is 0. This indicates that when $A_M^2$ is small, the empirical distribution $F_M(x)$ is considered close to the population distribution $F(x)$. As we have noted, we aim to focus on the tail of the distribution; it can be accomplished by Eq. (8). The one-sample AD test result $A_M^2$ can be used to verify if the given sample comes from a population with a specific distribution.

#### 2) Two-sample AD test

In our work, we want to verify that two sample sets come from the same unknown population. The two-sample AD test is suitable for this verification. The two-sample AD test [17], [18] is as follows. There are two sample empirical distributions $F_M(x)$ and $G_N(x)$. The $F_M(x)$ is an empirical distribution made from the set $\mathcal{F}$ with cardinality of samples set $M = |\mathcal{F}|$. The $G_N(x)$ is also an empirical distribution made from the set $\mathcal{G}$ with cardinality of samples set $N = |\mathcal{G}|$. $F_M(x)$ and $G_N(x)$ are the respective sample sets obtained independently from two different testing locations. The two-sample AD test can be used to verify if the both sample distributions come from the same distribution. In [17], [18], the two-sample version is defined as follows:

$$A_{MN}^2 = \frac{MN}{K} \int_{-\infty}^{\infty} \frac{(F_M(x) - G_N(x))^2}{H_K(x)(1 - H_K(x))} dH_K(x) \quad (9)$$

where $H_K(x) = (MF_M(x) + NG_N(x))/K$ with $K = M + N$. $A_{MN}^2$ is standardized to remove the dependencies derived by the number of samples. This standardized form is utilized to calculate the *p*-value [17], [18]. The *p*-value provides an evidence for hypotheses test.

The two-sample AD test is suitable to verify a hypothesis that two sample sets come from the same population. For the two-sample AD test, as a null hypothesis $H_0$, we set the $F_M(x)$ has the same population as $G_N(x)$. Also, we set that the $G_N(x)$ is an exponential distribution. Thus, if $F_M(x)$ and $G_N(x)$ comes from same population, namely $H_0$ is true, we may consider that $F_M(x)$ is the exponential distribution. If the *p*-value of AD test is large enough, it provides the evidence that $H_0$ is true.

The *p*-value is, under the assumption that the null hypothesis is true, the false-positive probability. A low *p*-value indicates that a test result provides evidence against the null hypothesis; a large *p*-value does not. Namely, large *p*-value denotes the probability of true-negative is low. The *p*-value is determined from the observation of the sample data. Thus, before observing the data, we set the threshold significance level(TSL), $TSL \in [0, 1]$, first. The TSL can be used to determine the critical value. Given a TSL and the number of samples that are used in the AD test, the TSL table in [18] is used to read off a value corresponding to the TSL and the number of samples. This read off value is called the critical value. If the standardized $A_{MN}^2$ is smaller than the critical value, this result indicates that the *p*-value is larger than the predefined TSL. In the TSL table of [18], the maximum TSL is 0.25. Thus, when standardized $A_{MN}^2$ is lower than the critical value corresponding to the 0.25 TSL, the *p*-value is capped at 0.25.

#### 3) Verification of the AD Test

In this subsection, we aim to verify the two-sample AD testing method. Verification is done under the assumption that the input distributions are *a priori* known. This will

**TABLE 1.** Example of the Anderson-Darling test results.

| The number of samples | Standardized $A_{MN}^2$ | $p$-value |
|---|---|---|
| 10 | -0.59 | $p \geq 0.250$ |
| 20 | 0.44 | $p = 0.21$ |
| 30 | 0.69 | $p = 0.17$ |

(a) $\mathcal{F} \sim \text{Exp}(1)$, $\mathcal{G} \sim \text{Normal}(1,1)$

| | | |
|---|---|---|
| 10 | 1.20 | $p = 0.103$ |
| 20 | 3.57 | $p = 0.011$ |
| 30 | 4.67 | $p = 0.004$ |

(b) $\mathcal{F} \sim \text{Exp}(1)$, $\mathcal{G} \sim \text{Exp}(2)$

| | | |
|---|---|---|
| 10 | 1.11 | $p = 0.113$ |
| 20 | -0.41 | $p \geq 0.250$ |
| 30 | -0.08 | $p \geq 0.250$ |

(c) $\mathcal{F} \sim \text{Exp}(1)$, $\mathcal{G} \sim \text{Exp}(1)$

**TABLE 2.** The observed frequency is calculated using the histogram in Fig. 4, and the expected frequency is calculated using the CDF of the exponential distribution derived from the Mean in Fig. 4

| Interval(%) | Observed frequency | Expected frequency |
|---|---|---|
| [0, 10) | 107 | 118.70 |
| [10, 20) | 82 | 71.73 |
| [20, 30) | 56 | 43.35 |
| [30, 40) | 20 | 26.19 |
| [40, 50) | 14 | 15.83 |
| [50, 60) | 7 | 9.56 |
| [60, 70) | 5 | 5.78 |
| [70, 80) | 4 | 3.49 |
| [80, 90) | 2 | 2.11 |
| [90, 100] | 3 | 1.27 |

clearly illustrate how we shall use the AD test and interpret its testing results.

In Table 1, we present three examples to give an insight into the $p$-value of the AD test; in this example, we use true distributions for $F_M(x)$ and $G_N(x)$. In Table 1, $\text{Exp}(\theta)$ indicates the exponential distribution with mean $\theta$ and $\text{Normal}(\mu, \sigma)$ indicates the normal distribution with mean $\mu$ and standard deviation $\sigma$. Namely, $\mathcal{F} \sim \text{Exp}(\theta)$ denotes that the sample set $\mathcal{F}$ of $F_M(x)$; samples are derived from the exponential distribution with mean $\theta$. In (a) of Table 1, we use the exponential distribution for $F_M(x)$ and the normal distribution for $G_N(x)$; these distributions have the same mean. This example presents, as the number of samples increases, the $p$-value tends to decrease if samples are drawn from different distributions. In (b) of Table 1, we set both $F_M(x)$ and $G_N(x)$ as the exponential distribution but each with different mean. This example presents, as the number of samples increases, even though samples are drawn from the same exponential distribution, the $p$-value tends to decrease if the means of distributions are different. In (c) of Table 1, we set both $F_M(x)$ and $G_N(x)$ as exactly the same exponential distribution. Namely, the two sample sets $\mathcal{F} \sim F_M(x)$ and $\mathcal{G} \sim G_N(x)$ come from the same population. This example shows, as the number of samples increases, that the $p$-value tends to increase when two sample sets are drawn from the same population. From these examples in Table 1, we note that, the closer the two distribution $F_M(x)$ and $G_N(x)$ are with each other, the larger $p$-value is obtained.

We aim to see whether the AD test result of our experiments indicates that $F_M(x)$ is close enough to $G_N(x)$. Namely, given there are two sample sets, one of $F_M(x)$ and the other of the exponential $G_N(x)$, we want to see if we can make a quality statement about how close the two sample sets are as a result of the AD test. The AD test result presents a significant $p$-value, i.e., $p \geq 0.25$; it is a necessary condition, but not a sufficient one, for the case that the two distributions are the same. In other words, if a decision of negating the null hypothesis is made, such that the distribution $F_M(x)$ is not close to $G_N(x)$ the exponential distribution, such a decision will incur an error with a probability greater than 0.25.

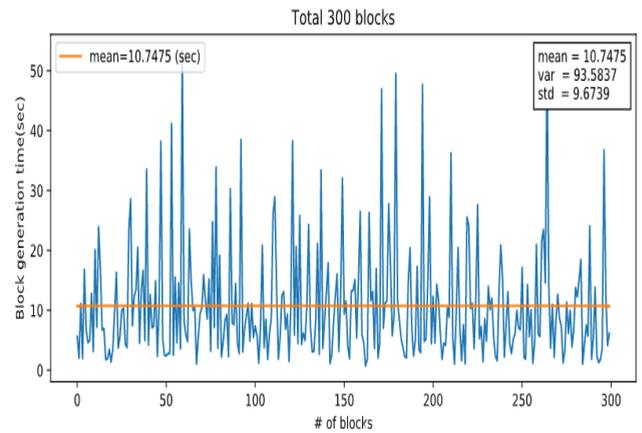

**FIGURE 4.** Plot of 300 BGTs when $n$ is 32. The legend at the top right shows the mean, variance, and standard deviation of BGT.

### D. Application of AD Test to BGT Distribution
In this subsection, we aim to apply the AD testing to determine the distribution of the BGT of ETH-ECC. For this experiment, 90 threads were used to generate a block. We experimented using a fixed code length to observe the BGT without difficulty change. In the test, two kinds of code length $n$ are used: 32 and 36. These are the two lowest types of code length $n$ in our pseudo-difficulty table used in the simulation. We divided the BGT into ten intervals between the minimum BGT and maximum BGT for a histogram. For example, when the minimum BGT is 10 and the maximum BGT is 20, there are ten intervals, i.e., [10,11], [11,12], ⋯, [19,20]. Using these intervals, we count the observed frequency of the BGT data. We set $F_M(x)$ using the observed frequency and set $G_N(x)$ using a mean of BGT data. For the expected frequency of $G_N(x)$ in Table 2, the mean in Fig. 4 is utilized. Namely, the mean in Fig. 4 is used as $1/\lambda$ for the CDF of the exponential distribution $G_N(x)$:

$$G_N(x) = 1 - e^{-\lambda x} \qquad (10)$$

TABLE 3. Anderson-Darling test result. The test result presents a large *p*-value. It denotes that if we reject the null hypothesis, the probability of true-negative is low.

| n | # of blocks | Observed mean(sec) | std | Standardized $A_{MN}^2$ | *p*-value |
|---|---|---|---|---|---|
| 32 | 100 | 10.86 | 9.84 | -1.12 | $p \geq 0.25$ |
| 32 | 200 | 11.24 | 10.16 | -1.20 | $p \geq 0.25$ |
| 32 | 300 | 10.74 | 9.67 | -1.18 | $p \geq 0.25$ |
| 32 | 400 | 11.08 | 9.84 | -1.09 | $p \geq 0.25$ |
| 32 | 500 | 10.91 | 9.62 | -1.11 | $p \geq 0.25$ |
| 32 | 600 | 10.87 | 9.48 | -0.80 | $p \geq 0.25$ |
| 32 | 700 | 10.84 | 9.41 | -0.36 | $p \geq 0.25$ |
| 32 | 800 | 10.76 | 9.40 | -0.36 | $p \geq 0.25$ |
| 36 | 100 | 56.00 | 55.20 | -1.11 | $p \geq 0.25$ |
| 36 | 200 | 51.04 | 49.71 | -1.19 | $p \geq 0.25$ |
| 36 | 300 | 47.84 | 45.49 | -1.12 | $p \geq 0.25$ |
| 36 | 400 | 49.97 | 47.80 | -1.19 | $p \geq 0.25$ |
| 36 | 500 | 49.24 | 46.95 | -1.11 | $p \geq 0.25$ |
| 36 | 600 | 48.23 | 46.96 | -1.18 | $p \geq 0.25$ |
| 36 | 700 | 48.36 | 47.68 | -1.18 | $p \geq 0.25$ |
| 36 | 800 | 48.03 | 46.89 | -1.18 | $p \geq 0.25$ |

The expected frequency of the Table 1 is calculated using the integral of $G_N(x)$ corresponding to the interval time. Because $G_N(x)$ is the exponential distribution, if $F_M(x)$ is close to $G_N(x)$, we may consider $F_M(x)$ is the exponential distribution.

### E. Discussion on AD Test Results

In Fig. 3, we present plots of the observed frequency and expected frequency. These frequencies are calculated in a manner mentioned in the subsection *Application of AD Test to BGT Distribution*. Fig. 3 shows that the observed frequency tends to follow the expected frequency. Also, in Table 3, the observed mean and standard deviation tend to converge as the number of blocks increase. Furthermore, in Table 3, we present the results of the AD test to discuss hypotheses $H_0$ and $H_a$. These results show a similar result of (c) in Table 3. In (c), we drew samples from the same true distribution; the results present the largest possible *p*-value. All the *p*-values in Table 3 have a larger than or equal to the 0.25 regardless of the number of blocks. In other words, if the null hypothesis is rejected, this decision will cause an error with a probability greater than 0.25. Namely, the decision that the BGT distribution $F_M(x)$ does not follows the exponential distribution could be made with a high decision error.

## VI. CONCLUSION

In this paper, we present the implementation, simulation, and validation of ETH-ECC. In the implementation, we showed how Ethereum applies ECCPoW as a consensus algorithm with real implementation. In the simulation, we conducted a multinode experiment using AWS EC2. The results revealed that the ECCPoW algorithm with varying difficulty is successfully implemented in the real world. In the validation, we showed the statistical results. These statistical results satisfy the necessary condition that the distribution of ECCPoW block generation time is exponential.